\documentclass[runningheads]{llncs}
\usepackage{cite}
\usepackage[colorlinks=true,linkcolor=blue,citecolor=blue,urlcolor=blue]{hyperref}
\usepackage{amsmath,amssymb,amsfonts}
\usepackage{algorithmic}
\usepackage{graphicx}
\usepackage{lipsum}
\usepackage{textcomp}
\usepackage{xcolor}
\usepackage{array}
\usepackage{cleveref}
\usepackage{caption}
\usepackage{subcaption}
\usepackage{arydshln}
\usepackage{tikz}
\usetikzlibrary{decorations.pathreplacing, positioning, arrows.meta}
\def\BibTeX{{\rm B\kern-.05em{\sc i\kern-.025em b}\kern-.08em
    T\kern-.1667em\lower.7ex\hbox{E}\kern-.125emX}}

\begin{document}
\title{A Unified Knowledge Graph to Permit Interoperability of Heterogeneous Digital Evidence}
%
%
\author{Ali Alshumrani\inst{1,2}\href{https://orcid.org/0000-0003-2078-2698}{0000-0003-2078-2698}\and
Nathan Clarke\inst{1}\href{https://orcid.org/0000-0002-3595-3800}{0000-0002-3595-3800} \and
Bogdan Ghita\inst{1}\href{https://orcid.org/0000-0002-1788-547X}{0000-0002-1788-547X}}
\authorrunning{Alshumrani et al.}
%
\institute{Centre for Cyber Security, Communications and Network Research (CSCAN), University of Plymouth, Plymouth, United Kingdom\\
\and
Department of Information Systems, Umm Al-Qura University, Makkah, Saudi Arabia\\
\email{\{ali.alshumrani, n.clarke, bogdan.ghita\}@plymouth.ac.uk}
}
\maketitle              
\begin{abstract}
The modern digital world is highly heterogeneous, encompassing a wide variety of communications, devices, and services. This interconnectedness generates, synchronises, stores, and presents digital information in multidimensional, complex formats, often fragmented across multiple sources. When linked to misuse, this digital information becomes vital digital evidence. Integrating and harmonising these diverse formats into a unified system is crucial for comprehensively understanding evidence and its relationships. However, existing approaches to date have faced challenges limiting investigators' ability to query heterogeneous evidence across large datasets. This paper presents a novel approach in the form of a modern unified data graph. The proposed approach aims to seamlessly integrate, harmonise, and unify evidence data, enabling cross-platform interoperability, efficient data queries, and improved digital investigation performance. To demonstrate its efficacy, a case study is conducted, highlighting the benefits of the proposed approach and showcasing its effectiveness in enabling the interoperability required for advanced analytics in digital investigations.

\end{abstract}
\section{Introduction}
The widespread use of digital devices and internet services has increased the
quantity and diversity of digital evidence \cite{vincze2016challenges}. This explosive growth has given rise to an unprecedented deluge of digital data, with daily volumes exceeding 2.5 quintillion bytes \cite{namjoshi2022role}. The surge in technology has not only transformed our daily lives but has also led to a corresponding surge in criminal activities, necessitating an increased demand for digital forensic services \cite{casey2019chequered}. Digital evidence, once primarily associated with cybercrime, now plays a pivotal role in investigating traditional criminal cases, with approximately 90\% of criminal investigations encompassing a digital footprint \cite{miller2023survey}.

However, the investigation of digital evidence has become increasingly challenging due to the sheer volume of data scattered across various evidence sources \cite{lillis2016current}. Each source has its unique file formats, structures, and schemes, creating a tapestry of heterogeneity and inconsistency in the nature of complex evidence. The complexity of integrating and unifying complex evidence effectively has significantly impacted digital forensics. The inherent heterogeneity and inconsistency in this data create a significant hurdle in achieving interoperability across digital sources \cite{rahman2020comprehensive}. Achieving interoperability across this diverse digital landscape necessitates the harmonisation and unification of these heterogeneous data types, a task that traditionally demands substantial effort and relies on leveraging a multitude of tools and methods \cite{rahman2020comprehensive, alshumrani2023unified, mohammed2018automating}. This need arises because traditional forensic tools, which are often tailored to specific technologies or platforms, tend to yield fragmented evidence. Consequently, in its isolated form, each piece of evidence requires manual examination and analysis to outline a comprehensive and correlated narrative \cite{alshumrani2023unified}. This manual intervention is time-consuming, error-prone, a cognitively taxing burden on investigators, and wholly inadequate to cope with the voluminous evidence encountered in contemporary digital environments \cite{casino2022research}.

In response to these challenges, the modern paradigm of evidence investigation requires a robust data harmonisation method capable of seamlessly integrating isolated evidence footprints into a unified system \cite{alshumrani2023unified}. Arguably, combining and representing heterogeneous data within a unified framework can address multiple challenges. These include tackling data heterogeneity and inconsistency, enhancing data automation, enabling advanced data correlation and visualisation across evidence resources, all while reducing the need for manual examination. This approach can also be instrumental in addressing other critical issues, such as anti-forensics techniques. For instance, unifying and cross-matching data from system-related events like web browser activity with network traffic logs can reveal whether a suspect is utilising some form of incognito function on the web browser to conceal their search history. Furthermore, the automated harmonisation, consolidation, and uniform structuring of evidence data can enhance the efficiency of digital investigations \cite{mohammed2018automating}.

Therefore, this study proposes the adoption of a Unified Metadata Graph Model (UMGM) as a solution to address the interoperability and harmonisation challenges posed by heterogeneous evidence. The proposed approach offers a standardised and unified methodology for representing fragmented and isolated evidence, regardless of its diverse sources or formats. By leveraging the power of graph database structures, this approach seamlessly integrates and harmonises evidence based on their associated metadata attributes, thereby enabling the potential for cross-platform interoperability within the realm of heterogeneous evidence. Furthermore, the adoption of the data graph database facilitates the unified knowledge representation of evidence data, capturing intricate relationships among entities, attributes, and events. Incorporating this proposed method empowers investigators with advanced query capabilities, interactive data refinement, and streamlined evidence analytics across the entire spectrum of evidence objects.

In the subsequent sections, Section~\ref{sec:lit} explores existing research on data ontologies and graphs and their applicability in achieving interoperability of heterogeneous digital evidence. Section~\ref{sec:sys} outlines the methodology and implementation of the unified system architecture, detailing its functionalities and features. In Section~\ref{sec:uc}, a hypothetical case study is conducted to illustrate the effectiveness of the proposed approach. Finally, in Section~\ref{sec:con}, the findings are summarised, and promising directions for future research are outlined.

\section{Literature Review}
\label{sec:lit}
Data ontology holds a pivotal position within the semantic web, offering a formalised and standardised definition of concepts, relationships, and properties that characterise evidence-related information. The adoption of data ontology within the realm of digital evidence has been the subject of many studies. Authors in~\cite{brady2015deso} developed a Digital Evidence Semantic Ontology (DESO) to index and classify evidence artefacts, focusing primarily on their discovery locations. To gauge the effectiveness of DESO, they undertook a case study that examined evidence from two distinct computer systems and USB memory devices. The findings from this study indicated that the approach mainly identified artefacts by their location, overlooking related properties. Moreover, the technique employed for data correlation based on relevance was not adequately detailed, leading to the method's suboptimal utilisation, as it was heavily reliant on the expertise of the digital examiner. In a related study, \cite{chabot2015ontology} introduced the Semantic Analysis of Digital Forensic Cases (SADFC) system. This system was designed to reconstruct and analyse timelines relevant to evidence incidents. It employs the Ontology for the Representation of Digital Incidents and Investigations (ORD2I) as a formal logical language, capturing the semantics of concepts and relationships within digital evidence and thereby facilitating knowledge representation. To assess the proposed approach, the authors conducted experiments with an experimental malware dataset, which included browser-related data, such as browsing histories and downloaded files. The experiments showcased SADFC's ability to effectively extract knowledge from file system artefacts, conduct data analysis, and address queries. However, it is noteworthy that SADFC faces challenges in the seamless integration, analysis, and validation of complex evidence, especially when dealing with a vast volume of heterogeneous data without human intervention.

Seeking to standardise the representation of data objects and relationships to enhance correlation among evidence sources, \cite{casey2015leveraging} leveraged the capabilities of the Cyber Observable eXpression (CybOX) schema and the Unified Cyber Ontology (UCO) to develop an ontology named Digital Forensic Analysis eXpression (DFAX). DFAX enhances CybOX's functionalities by offering a more detailed depiction of forensic-relevant information, covering activities executed by both subjects and forensic examiners. It also incorporates UCO's general abstractions to represent concepts that span the cyber domain, thereby facilitating more advanced forensic analysis. This initiative laid the groundwork for the introduction of the Cyber-investigation Analysis Standard Expression (CASE) ontology \cite{casey2018evolution}. Developed in collaboration with the UCO ontology, CASE consistently represents constructs across various cyber-centric domains, enhancing interoperability among different evidence domains. However, while these systems primarily focus on consolidating and formatting evidence data to facilitate its exchange across a broad spectrum of cyber-related domains, they do not address the essential need to harmonise evidence data in a standardised and unified manner. Such harmonisation would enable the application of advanced data correlation methods and more effective searching of the evidence data.

In research conducted in \cite{arshad2020formal}, introduced an Event-based Forensic Integration Ontology for Online Social Networks (EFIOSN) as a formal knowledge model for constructing and automating evidence analysis. The application of this model was demonstrated through a theoretical case study involving a defamation attempt on the Twitter platform. The study utilised timeline analysis based on temporal activity patterns to establish the sequential order of evidence-related events. The findings revealed that EFIOSN successfully identified associations and similarities among potential events, serving as an initial step for further investigation and forming direct evidence. However, it is important to note that the proposed model assumes the ontology’s adequacy in capturing all pertinent aspects of online social network data, indicating that the completeness and comprehensiveness of the ontology may not be sufficient. In a further effort to semi-automate network packet analysis, \cite{sikos2020knowledge} proposed a Packet Analysis Ontology (PAO) with the intention of providing a formal representation of concepts and properties related to packet analysis. To assess the effectiveness of this ontology, a case study was conducted using honeypot data, which simulates critical infrastructure containing Supervisory Control and Data Acquisition (SCADA) components. The ontology was applied to capture data semantics based on Wireshark frames, including packet frame numbers, timestamps, source and destination IP addresses, protocol numbers, and frame length values. The results demonstrated that the proposed system achieved a broader concept and role semantics range compared to similar studies with similar objectives. However, inherent limitations of this approach include its restriction to capturing data aligned with Wireshark frames only and its scalability mechanisms for validating, interacting with, and refining the captured data. These limitations may impede the ability to perform complex analyses and obtain desired results.

In a subsequent study, \cite{chikul2021ontology} introduced ForensicFlow system. This system built upon the Web Ontology Language (OWL), a semantic web technology, is designed to facilitate knowledge integration and enable semantic querying of data relationships using a query language known as SPARQL Protocol and Resource Description Framework (RDF) Query Language (SPARQL). For their experimental setup, the researchers conducted a case study focused on a ransomware attack scenario. This scenario involved various digital artefacts, including memory dumps and a disk image of a Windows operating system. The proposed ontology was instantiated by analysing these artefacts and querying Windows Prefetch data using SPARQL. This process successfully revealed two suspicious events and their dependencies. However, it's worth noting that the system's scope is confined to specific high-level events within the operating system. This limitation may hinder its applicability in more intricate cases. Additionally, the system lacks automated data validation capabilities, which could affect its robustness when handling large volumes of related data.

In the context of utilising data graph databases to automate digital forensic analyses, \cite{noel2016cygraph} proposed a CyGraph system for capturing data relationships across various network and host source entities, thereby facilitating cybersecurity analytics and the visual representation of security events.  The efficacy of this system was assessed via a case study, which encompassed an attack scenario executed on several internally vulnerable host systems. This case study underscored the system's proficiency in executing data queries and visually mapping potential attack patterns. However, the system's scalability is primarily confined to the predefined data graph model, which is predominantly designed for network-oriented environments. In a parallel vein, \cite{schelkoph2019digital} proposed a Property Graph Event Reconstruction (PGER) system with an emphasis on data normalisation and correlation. During the evaluation, researchers applied the PGER to a sample dataset containing events from web browsing, downloaded files from various web browser applications, document application events, and other system-related events. The system demonstrated the ability to index traversal and discover adjacent node entities to some extent, defining high-level rules through a combination of terms. However, it did present limitations in providing a holistic abstraction, given that the expert rules formulated within the system did not encapsulate all system events.

Whilst the aforementioned studies undeniably provide pivotal advancements in the field, there remain areas that necessitate further improvements. The scalability of current methodologies requires enhancement, particularly in terms of harmonising diverse data types, refining data relationships, optimising intelligent data querying mechanisms, and fostering sophisticated cross-data analytics. A considerable limitation of ontological studies lies in their circumscribed scope, which solely maps data relationships pertinent to specific technological objects. This constraint implies that such methodologies are relegated to identifying and extracting data that aligns with their predefined classifications. Ontological studies frequently concentrate on discrete subsets of digital traces, whilst failing to encompass the full spectrum of relevant evidence information. Their predominant reliance on specific types of data analysis, such as timeline analysis, may inadvertently introduce constraints, potentially limiting the effectiveness and scalability of investigations. Conversely, studies employing graph-based approaches demonstrate the potential to capture data relationships and perform data queries, yet they encounter limitations concerning the level of abstraction and interpretability of heterogeneous events. While these approaches have enhanced data model scalability and query efficiency, further research is crucial to devise a comprehensive solution addressing the challenges of data heterogeneity and interoperability within a consistent, interactive framework. The development of more advanced and robust tools that adeptly integrate, harmonise, and analyse intricate relationships within heterogeneous evidence will enable investigators to query, navigate, refine, and define the evidence more efficiently, leading to more effective forensic investigations.

\section{Unified Metadata Graph Model}
\label{sec:sys}
\begin{figure}
    \centering
    \includegraphics[width=1\linewidth]{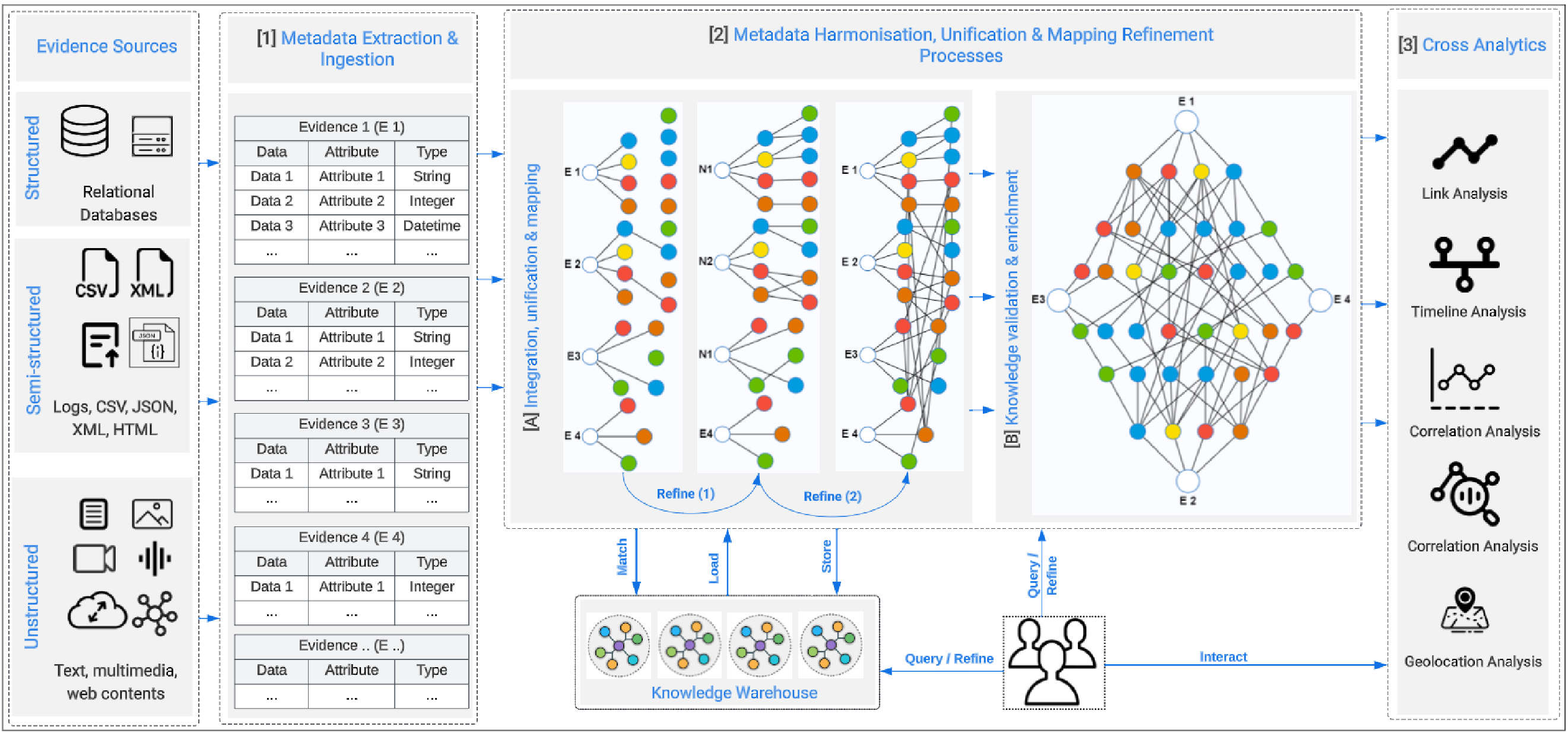}
    \vspace{-18.5pt} 
    \caption{A generic workflow of the unified metadata graph model.}
    \label{fig1}
\end{figure}

The UMGM introduces a structured framework, divided into three pivotal phases, that is aimed at effectively integrating, harmonising, and unifying diverse data objects. The first phase lays the foundation, focusing on the extraction and ingestion of evidence metadata, ensuring all essential data is captured and ready for processing. The second phase delves into metadata harmonisation and unification, which is further categorised into two subphases: metadata integration and mapping; and knowledge validation and enrichment. The third phase shifts the focus towards extracting actionable insights, concentrating on conducting advanced analytics across the resultant homogeneous data, aiding investigators in discerning significant patterns and conclusions. Fig.~\ref{fig1} provides a general overview of the system's workflow process, demonstrating a visual representation of the methodological progression. The subsequent sections will delve deeper, elucidating the objectives associated with each phase within the UMGM framework.

\subsection{Metadata  Extraction and Ingestion}
The metadata extraction and ingestion process is designed to capture and incorporate relevant attributes from various forms of evidence. This evidence encompasses structured data from relational databases, semi-structured formats such as logs, CSV, JSON, XML, and HTML from web applications and configuration files, as well as unstructured data including text documents, multimedia files, and dynamic web content. Given the extensive range of data sources, the resulting metadata can represent a myriad of types. These may consist of attributes like usernames, timestamps, file sizes, geolocation data, MIME types, email IDs, phone numbers, IP addresses, and network protocols, among other significant properties. This phase subsequently processes these multi-dimensional attributes, transmitting and adapting them into the system for subsequent analyses and operations.

\subsection{Metadata Harmonisation, Unification, and Mapping Refinement Processes}
During this phase, the extracted metadata attributes are defined, integrated, and transformed into a standardised, unified graph database format. This transformation ensures the harmonisation of metadata from various data sources into a consistent structure, highlighting the advantages of using a graph database. The main components of this unified data graph structure include nodes, properties, and relationships. Nodes represent individual data entities, with each node corresponding to specific metadata that provides detailed insights about the associated data properties. Properties offer further specific information related to each node. Relationships define the connections, associations, or dependencies among nodes, capturing the intricate structure and semantics between data entities. Presenting data in this manner facilitates a holistic understanding of evidence attributes, their characteristics, and relationships. This structure notably enables efficient cross-querying and exploration of evidence connections. As a result, investigators gain a comprehensive view of metadata-associated properties from multiple evidence data sources. They can also effectively utilise cross-search functionalities to retrieve and refine evidence attribute relationships. These objectives are achieved through two specific sub-phases, which will be outlined in the subsequent sections. 

\subsubsection{Unification, Mapping, and Refinement Process:}

This process harmonises heterogeneous evidence efficiently. It standardises and aligns metadata attributes using predefined data model characteristics from the knowledge warehouse. Investigators then actively engage with evidence entities and relationships, which sheds light on interdependencies and associations. The unified system retrieves metadata attributes from extracted and ingested metadata, and this raw data undergoes automatic refinement based on predefined criteria matched with the knowledge warehouse. The first refinement stage, represented as 'Refine 1', establishes initial relationships between instances, enhancing data coherence and structure. The refined metadata is then made available to investigators for further enhancement. In the second refinement stage, denoted as 'Refine 2', investigators identify and establish new relationships between instances, as depicted in Fig.~\ref{fig1}. This iterative process of refining and enhancing data enables investigators to gain insights into the interconnectedness and dependencies among entities, thus supporting a more comprehensive and accurate digital investigation. It addresses discrepancies and conflicts in metadata, promoting data interoperability and consistency. Additionally, it reveals connections across diverse evidence sources, offering a panoramic evidence perspective.

\subsubsection{Knowledge Validation and Enrichment:}
The knowledge validation and enrichment phase builds upon previously refined metadata relationships. It further involves human interaction and validation mechanisms to ensure the accuracy and reliability of the generated models of evidence metadata. Investigators actively engage in this phase to validate and identify areas for potential enhancement, thereby further enriching the evidence. By leveraging the defined data relationships, investigators retrieve additional relevant data based on the newly established connections. Moreover, investigators verify the accuracy, completeness, and reliability of the harmonised metadata through this knowledge validation process, ensuring the integrity of the evidence and its associated relationships. Knowledge enrichment plays a vital role in this phase, involving the integration of additional contextual or domain-specific knowledge into the metadata. This process enhances the richness and implications of the evidence, providing a holistic perspective on related evidence and a more comprehensive understanding of the case. This ensures a well-informed approach before proceeding with cross-data analytics, making the digital forensics investigation more effective and accurate. 

\subsection{Cross Evidence Analytics}
The cross-analytic phase equips investigators with the capability to apply various evidence-analytical techniques to the harmonised and enriched knowledge graph, aiding in identifying patterns and correlations across evidence sources. This facilitates the determination of critical evidence. The system's interactive data exploration functionalities allow for the retrieval and visual representation of predefined data models. This enables investigators to review, exclude or focus on specific nodes, add information, and update changes for further analysis. Specifically, the system supports Link Analysis, identifying relationships between entities; Timeline Analysis, showcasing events chronologically; Correlation Analysis, assessing relationships among data points; and Geolocation Analysis, analysing the geographical data associated with evidence. Moreover, built-in queries offer searches based on attributes such as username, timeline frame, geolocation, email ID, IP address, and keyword search. The system also suggests visualisation types and allows combining methods within one interface, enhancing evidence analysis. Advanced analytics enhance the efficiency of forensic investigations. The system's capabilities facilitate a comprehensive exploration of evidence data, bolstering overall investigative outcomes.

\section{Use Case Evaluation} \label{sec:uc}
\label{sec:us}
This section presents a hypothetical data leakage case study that showcases the practical application and capabilities of the proposed system. The aim is to demonstrate how the unified knowledge graph model can formalise and streamline the investigation process by seamlessly integrating and analysing evidence from diverse sources. The case study revolves around an intricate investigation scenario that involves multiple sources of evidence. Through the application of the unified knowledge graph model, this illustrative case study highlights the effectiveness and efficiency of the approach in addressing the challenges posed by heterogeneous evidence in digital forensics investigations.

\subsection{Crime Scenario Overview}
\textit{Corporation AIxz specialises in developing Artificial Intelligence (AI) models for financial institutions. These highly valuable models are securely stored in a dedicated cloud server, which can only be accessed through a limited list of static IP addresses within the corporation's local network. However, on May 19\textsuperscript{th} 2022, an incident of data leakage occurred involving an unauthorised transfer of a highly confidential AI model to an external source. In response to this breach, the corporation's digital investigation team swiftly initiated an initial examination of the internal network to determine the source of the breach. During the initial investigation, the team identified a high volume of encrypted traffic originating from various internal IP addresses. Thus, based on this determination, they uncovered compelling evidence implicating several systems connected to employees named Alex, Bill, Lisa, and Abby, as well as an unidentified IP address associated with an unrecognised host system. Recognising the seriousness of the situation, the team promptly seized several digital resources, as indicated in Table~\ref{tab:table1}, \textit for further analysis and investigation to identify the insider(s) responsible for the breach.}

\begin{table}[htbp]
\caption{Summary of the seized digital sources.}
\label{tab:table1}
\resizebox{1\textwidth}{!}{%
\begin{tabular}{l|l}
\hline
Evidence & Description \\
\hline
Network logs & An initial collection of network traffic logs covering the past 30 days \\
\hline
Computers & Several Windows operating systems \\
\hline
Memory artefacts & A collection of computers' volatile memory artefacts \\
\hline
Servers & Several internal server logs, including Syslog, Email, DNS, and web application servers \\
\hline
Cloud storage & A dedicated cloud storage logs for storing the developed AI models \\
\hline
Firewall logs & An initial collection of perimeter firewall logs covering the past 30 days \\
\hline
\end{tabular}%
}
\end{table}

\subsection{Investigation Objectives}
The primary objective of this investigation was to analyse the seized evidence and determine the responsibility of the identified suspects for the leakage of the AI model. Consequently, the following investigative questions were postulated:
    \begin{enumerate}
        \item Investigate whether the identified suspects were involved in transmitting the AI model.
        \label{Q1}
        \item Investigate the origin of the unidentified IP and analyse patterns and anomalies across the seized evidence.
        \label{Q2}
        \item Conduct link analysis across all potential data objects and identify relationships among the identified pieces of evidence.
        \label{Q3}
        \item Based on data harmonisation results, perform timeline visualisation analysis and highlight the potential patterns or anomalies of evidence events chronologically.
        \label{Q4}
    \end{enumerate}
\subsection{Investigation Workflow}
This subsection outlines the sequential phases of the proposed system in investigating the data leakage case study.
\subsubsection{Metadata Extraction:}

In the context of the seized evidence sources mentioned in Table~\ref{tab:table1}, a crucial step in the proposed system's investigation process is the identification and extraction of the relevant metadata attributes. Table \ref{tab:table2}  provides illustrative examples of these evidence metadata attributes, encompassing memory, network logs, cloud data, and Syslog attributes. These metadata attributes correspond to the extracted artefacts that hold the utmost relevance to the investigation.

\begin{table}[htbp]
\caption{A simplified example of evidence metadata attributes.}
\label{tab:table2}
\small  
\begin{subtable}{1\textwidth}
\caption{Sample memory artefacts.}
\vspace{-5pt} 
\label{tab:table21}
\resizebox{\textwidth}{!}{%
\begin{tabular}{l|l|l|l|l|l|l}
\hline
Process Name & Protocol & Local IP Address & Foreign Address & State & CPU Time & Elapsed Time \\
\hline
Putty & TCP & 10.0.0.20:52814 & 10.0.0.100:ssh & Established & 0:00:11.812 & 0:08:58.851 \\
\hline
\end{tabular}%
}
\end{subtable}
\begin{subtable}{1\textwidth}
\caption{Sample network PCAP logs.}
\vspace{-5pt} 
\label{tab:table3} 
\resizebox{\textwidth}{!}{%
\begin{tabular}{l|l|l|l|l|l|l|l|l} \hline  
Timestamp & Source MAC & Destination MAC & Source IP & Destination IP & Source Port & Destination Port & Protocol & Host \\ \hline
18/5/2022 10:10:05 & Ff:df:f9:c4:94:ac & 24:d4:4b:8e:02:86 & 10.0.0.20 & 10.0.0.100 & 52814 & 22 & SSHv2 & System1 \\ \hline
18/5/2022 10:10:40 & Ff:df:f9:c4:94:ac & 24:d4:4b:8e:02:86 & 10.0.0.20 & 10.0.0.100 & 49130 & 443 & HTTPS & System1 \\ \hline
18/5/2022 10:55:12 & 98:0c:b9:99:3f:5b &   ---  & 10.0.0.15 & 93.125.188.220 & 49190 & 443 & HTTPS &     ---    \\ \hline         
\end{tabular}%
}
\end{subtable}
\begin{subtable}{1\textwidth}
\caption{Sample cloud metadata attributes.}
\vspace{-5pt} 
\label{tab:table4}
\resizebox{\textwidth}{!}{%
\begin{tabular}{l|l|l|l|l|l} \hline  
File Name & File Size & Created By & Created Timestamp & Accessed By & Accessed Timestamp \\ \hline
FinAI.h5 & 4.2 GB & AIxz & 20/04/2022 13:20:22 & Alex@AIxz.ai & 18/5/2022 10:10:40 \\ \hline       
\end{tabular}%
}
\end{subtable}
\begin{subtable}{1\textwidth}
\caption{Sample syslog server events.}
\vspace{-5pt} 
\label{tab:table5}
\resizebox{\textwidth}{!}{%
\begin{tabular}{l|l|l|l|l|l} \hline
Device Name & IP Address & Event Type & Created Timestamp & Accessed By & Accessed Timestamp \\ \hline
Perimeter firewall & 10.0.0.200 & Logon & 20/04/2022 13:20:22 & Alex & 18/5/2022 10:20:20 \\ \hline  
\end{tabular}%
}
\end{subtable}
\end{table}

\noindent
Table \ref{tab:table21} provides a sample of memory logs detailing a network connection established using a “Putty” application. These logs were generated from a host system with an IP address of 10.0.0.20 via port number 52814. They include information about the connection state, CPU time, and elapsed time. In the context of the collected network PCAP (Packet Capture), Table~\ref{tab:table3} shows a sample view of the extracted logs containing timestamps, MAC and IP addresses, ports, protocol, and the hosts involved in the communication. It provides information about network traffic events and their corresponding details. Furthermore, Table~\ref{tab:table4} provides information about cloud sample events. It indicates the creation timestamps of the model, includes the model named “FinAI.h5”, and specifies its size as 4.2 GB. Finally, Table~\ref{tab:table5} showcases sample events captured by the Syslog server, highlighting attributes such as device name, event type, creation timestamp, and the individual who accessed the device. These logs refer to a login attempt on the perimeter firewall, specifying the source IP address, accessing user name, and the relevant timestamps.

Notably, the examination of these multitudes of metadata attributes using traditional methodologies necessitates investigators to transition between an array of tools and techniques each tailored to specific data types. For instance, when analysing the metadata illustrated in ~\Cref{tab:table21,tab:table3,tab:table4,tab:table5} investigative teams would typically employ distinct specialised tools and manual methodologies for each data source. They commonly employ tools like Volatility or Rekall for memory dump analysis and Wireshark or TCPdump for inspecting network packets. Furthermore, they might manually cross-reference access logs or employ dedicated tools to investigate event logs found in Syslog and cloud data records. Whilst the unified knowledge graph seeks to thoroughly extract, examine, and harmonise diverse metadata types into a unified graph database, thus enabling interactive refinement by investigators and ensuring compatibility with heterogeneous digital evidence. Through the proposed approach, investigators can explore interconnected metadata attributes, effectively identifying correlations and patterns that may have been overlooked in traditional investigations. Furthermore, the graph's query and refinement features provide a more streamlined and intuitive method for extracting information, reducing the reliance on disparate tools and manual efforts. The following subsection will demonstrate how these diverse metadata attributes are effectively harmonised and mapped, highlighting the efficiency of the proposed model in facilitating comprehensive digital evidence investigations.

\subsubsection{Metadata Mapping Process:}
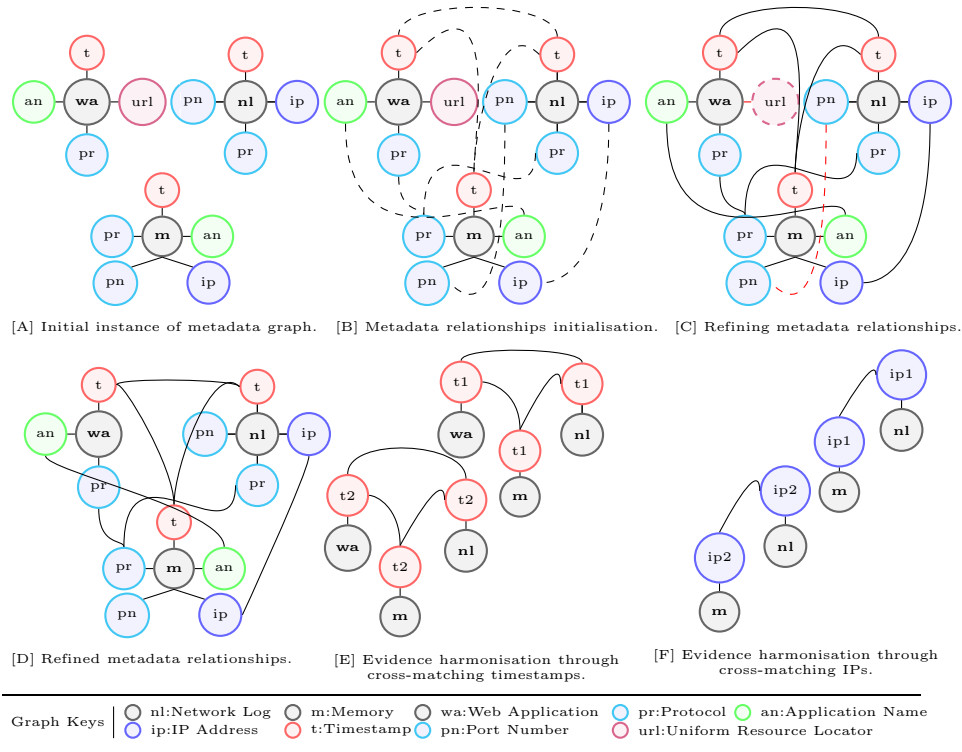
\begin{figure}[htbp]
\centering
\begin{minipage}{0.35\textwidth}
    \begin{tikzpicture}[
        AA/.style={circle, draw=black!60, fill=black!5, thick, minimum size=0.01cm, font=\tiny},
        BB/.style={circle, draw=red!60, fill=red!5, thick, minimum size=0.01cm, font=\tiny},
        CC/.style={circle, draw=blue!60, fill=blue!5, thick, minimum size=0.01cm, font=\tiny},
        DD/.style={circle, draw=green!60, fill=green!5, thick, minimum size=0.01cm, font=\tiny},
        EE/.style={circle, draw=yellow!60, fill=yellow!5, thick, minimum size=0.01cm, font=\tiny},
        FF/.style={circle, draw=orange!60, fill=orange!5, thick, minimum size=0.01cm, font=\tiny},
        GG/.style={circle, draw=cyan!60, fill=blue!5, thick, minimum size=0.01cm, font=\tiny},
        QQ/.style={circle, draw=purple!60, fill=purple!5, thick, minimum size=0.01cm, font=\tiny},
        ]
        \node at (-1.58, -3) [font=\tiny] {\tiny[{A}] Initial instance of metadata graph.};
        \node [AA, xshift=-0.5cm] (net) {\textbf{nl}};
        \node[CC] (sipnet) [right=.1cm of net] {\tiny ip};
        \node[GG] (prnet) [below=.1cm of net] {\tiny pr};
        \node[BB] (tsnet) [above=.1cm of net] {\tiny t};
        \node[GG] (pnnet) [left=.1cm  of net] {\tiny pn};
        \draw[-, thin] (net.east) to (sipnet.west);
        \draw[-, thin] (net.north) to (tsnet.south);
        \draw[-, thin] (net.south) to (prnet.north);
        \draw[-, thin] (net.west) to (pnnet.east);
        \node [AA] (m)         [below left=1 of prnet]            {\textbf{m}};
        \node[BB] (tsm)        [above=.1 of m]                        {t};
        \node[DD]  (anm)       [right=.1 of m]                        {an}; 
        \node[GG] (prm)         [left=.1 of m]                        {pr};
        \node[CC] (sipm)       [below right=.3 of m]                  {ip};
        \node[GG] (pnm)       [below left=.3 of m]                    {pn};
        \node [AA] (wa)        [left=1.5 of net]                  {\textbf{wa}};
        \node[BB] (twa)        [above=.1 of wa]                       {t};
        \node[QQ] (urlwa)       [right=.1 of wa]                      {url}; 
        \node[DD] (anwa)        [left=.1 of wa]                       {an};
        \node[GG] (prwa)        [below=.1 of wa]                      {pr};
        \draw[-, thin] (net.east) to (sipnet.west);
        \draw[-, thin] (net.north) to (tsnet.south);
        \draw[-, thin] (net.south) to (prnet.north);
        \draw[-, thin] (net.west) to (pnnet.east);
        \draw[-, thin] (m.east) to (anm.west);
        \draw[-, thin] (m.north) to (tsm.south);
        \draw[-, thin] (m.south) to (sipm.north west);
        \draw[-, thin] (m.south) to (pnm.north east);
        \draw[-, thin] (m.west) to (prm.east);
        \draw[-, thin] (wa.east) to (urlwa.west);
        \draw[-, thin] (wa.north) to (twa.south);
        \draw[-, thin] (wa.south) to (prwa.north);
        \draw[-, thin] (wa.west) to (anwa.east);
        \draw[dashed, -, thin, white] (twa.east) .. controls  +(north east:16mm) and +(north east:6mm)   .. (tsm.north);
        \end{tikzpicture}
    \end{minipage}%
\begin{minipage}{0.35\textwidth}
    \begin{tikzpicture}[
       AA/.style={circle, draw=black!60, fill=black!5, thick, minimum size=0.01cm, font=\tiny},
        BB/.style={circle, draw=red!60, fill=red!5, thick, minimum size=0.01cm, font=\tiny},
        CC/.style={circle, draw=blue!60, fill=blue!5, thick, minimum size=0.01cm, font=\tiny},
        DD/.style={circle, draw=green!60, fill=green!5, thick, minimum size=0.01cm, font=\tiny},
        EE/.style={circle, draw=yellow!60, fill=yellow!5, thick, minimum size=0.01cm, font=\tiny},
        FF/.style={circle, draw=orange!60, fill=orange!5, thick, minimum size=0.01cm, font=\tiny},
        GG/.style={circle, draw=cyan!60, fill=blue!5, thick, minimum size=0.01cm, font=\tiny},
        QQ/.style={circle, draw=purple!60, fill=purple!5, thick, minimum size=0.01cm, font=\tiny},
        ]
        \node at (-5.8, -3) [font=\tiny] {\tiny[{B}] Metadata relationships initialisation.};
        \node [AA, xshift=-5cm] (net) {\textbf{nl}};
        \node[CC] (sipnet) [right=.1cm of net] {\tiny ip};
        \node[GG] (prnet) [below=.1cm of net] {\tiny pr};
        \node[BB] (tsnet) [above=.1cm of net] {\tiny t};
        \node[GG] (pnnet) [left=.1cm  of net] {\tiny pn};
    
        \draw[-, thin] (net.east) to (sipnet.west);
        \draw[-, thin] (net.north) to (tsnet.south);
        \draw[-, thin] (net.south) to (prnet.north);
        \draw[-, thin] (net.west) to (pnnet.east);
        \node [AA] (m)         [below left=1 of prnet]            {\textbf{m}};
        \node[BB] (tsm)        [above=.1 of m]                          {t};
        \node[DD]  (anm)       [right=.1 of m]                          {an}; 
        \node[GG] (prm)        [left=.1 of m]                           {pr};
        \node[CC] (sipm)       [below right=.3 of m]                    {ip};
        \node[GG] (pnm)        [below left=.3 of m]                     {pn};
        \node [AA] (wa)        [left=1.5 of net]            {\textbf{wa}};
        \node[BB] (twa)        [above=.1 of wa]                        {t};
        \node[QQ] (urlwa)      [right=.1 of wa]                      {url}; 
        \node[DD] (anwa)       [left=.1 of wa]                       {an};
        \node[GG] (prwa)       [below=.1 of wa]                      {pr};
        \draw[-, thin] (net.east) to (sipnet.west);
        \draw[-, thin] (net.north) to (tsnet.south);
        \draw[-, thin] (net.south) to (prnet.north);
        \draw[-, thin] (net.west) to (pnnet.east);
        \draw[-, thin] (m.east) to (anm.west);
        \draw[-, thin] (m.north) to (tsm.south);
        \draw[-, thin] (m.south) to (sipm.north west);
        \draw[-, thin] (m.south) to (pnm.north east);
        \draw[-, thin] (m.west) to (prm.east);
        \draw[-, thin] (wa.east) to (urlwa.west);
        \draw[-, thin] (wa.north) to (twa.south);
        \draw[-, thin] (wa.south) to (prwa.north);
        \draw[-, thin] (wa.west) to (anwa.east);
        \draw[dashed, -, thin, black] (tsnet.north) .. controls  +(north:5mm) and +(north:5mm)   .. (twa.north);
        \draw[dashed, -, thin, black] (tsnet.west) .. controls  +(north west:8mm) and +(north:5mm)   .. (tsm.north);
        \draw[dashed, -, thin, black] (twa.east) .. controls  +(north east:16mm) and +(north:5mm)   .. (tsm.north);
        \draw[dashed, -, thin, black] (anwa.south) .. controls  +(south:22mm) and +(north:5mm)   .. (anm.north);
        \draw[dashed, -, thin, black] (sipnet.south) .. controls  +(south:22mm) and +(north:0mm)   .. (sipm.east);
        \draw[dashed, -, thin, black] (prnet.west) .. controls  +(south:10mm) and +(north:15mm)   .. (prm.north);
        \draw[dashed, -, thin, black] (prwa.south) .. controls  +(south:5mm) and +(north:1mm)          .. (prm.north);
        \draw[dashed, -, thin, black] (pnnet.south) .. controls  +(south:8mm) and +(south east:10mm)   .. (pnm.east);
        \end{tikzpicture}
    \end{minipage}%
\begin{minipage}{0.35\textwidth}
    \begin{tikzpicture}[
        AA/.style={circle, draw=black!60, fill=black!5, thick, minimum size=0.01cm, font=\tiny},
        BB/.style={circle, draw=red!60, fill=red!5, thick, minimum size=0.01cm, font=\tiny},
        CC/.style={circle, draw=blue!60, fill=blue!5, thick, minimum size=0.01cm, font=\tiny},
        DD/.style={circle, draw=green!60, fill=green!5, thick, minimum size=0.01cm, font=\tiny},
        EE/.style={circle, draw=yellow!60, fill=yellow!5, thick, minimum size=0.01cm, font=\tiny},
        FF/.style={circle, draw=orange!60, fill=orange!5, thick, minimum size=0.01cm, font=\tiny},
        GG/.style={circle, draw=cyan!60, fill=blue!5, thick, minimum size=0.01cm, font=\tiny},
        QQ/.style={circle, draw=purple!60, fill=purple!5, thick, minimum size=0.01cm, font=\tiny},
        ]
        \node at (-6.8, -3) [font=\tiny] {\tiny[{C}] Refining metadata relationships.};
        \node [AA, xshift=-6cm] (net) {\textbf{nl}};
        \node[CC] (sipnet) [right=.1cm of net] {\tiny ip};
        \node[GG] (prnet) [below=.1cm of net] {\tiny pr};
        \node[BB] (tsnet) [above=.1cm of net] {\tiny t};
        \node[GG] (pnnet) [left=.1cm  of net] {\tiny pn};
        \draw[-, thin] (net.east) to (sipnet.west);
        \draw[-, thin] (net.north) to (tsnet.south);
        \draw[-, thin] (net.south) to (prnet.north);
        \draw[-, thin] (net.west) to (pnnet.east);
        \node [AA] (m)         [below left=1 of prnet]            {\textbf{m}};
        \node[BB] (tsm)        [above=.1 of m]                       {t};
        \node[DD]  (anm)       [right=.1 of m]                      {an}; 
        \node[GG] (prm)         [left=.1 of m]                       {pr};
        \node[CC] (sipm)       [below right=.3 of m]                      {ip};
        \node[GG] (pnm)       [below left=.3 of m]                      {pn};
        \node [AA] (wa)        [left=1.5 of net]            {\textbf{wa}};
        \node[BB] (twa)        [above=.1 of wa]                       {t};
        \node[QQ, dashed] (urlwa)       [right=.1 of wa]               {url}; 
        \node[DD] (anwa)        [left=.1 of wa]                       {an};
        \node[GG] (prwa)        [below=.1 of wa]                     {pr};
        \draw[-, thin] (net.east) to (sipnet.west);
        \draw[-, thin] (net.north) to (tsnet.south);
        \draw[-, thin] (net.south) to (prnet.north);
        \draw[-, thin] (net.west) to (pnnet.east);
        \draw[-, thin] (m.east) to (anm.west);
        \draw[-, thin] (m.north) to (tsm.south);
        \draw[-, thin] (m.south) to (sipm.north west);
        \draw[-, thin] (m.south) to (pnm.north east);
        \draw[-, thin] (m.west) to (prm.east);
        \draw[-, thin, dashed, red] (wa.east) to (urlwa.west);
        \draw[-, thin] (wa.north) to (twa.south);
        \draw[-, thin] (wa.south) to (prwa.north);
        \draw[-, thin] (wa.west) to (anwa.east);
        \draw[-, thin, black] (tsnet.north) .. controls  +(north:5mm) and +(north:5mm)   .. (twa.north);
        \draw[-, thin, black] (tsnet.west) .. controls  +(north west:8mm) and +(north:5mm)   .. (tsm.north);
        \draw[-, thin, black] (twa.east) .. controls  +(north east:16mm) and +(north:5mm)   .. (tsm.north);
        \draw[-, thin, black] (anwa.south) .. controls  +(south:22mm) and +(north:5mm)   .. (anm.north);
        \draw[-, thin, black] (sipnet.south) .. controls  +(south:22mm) and +(north:0mm)   .. (sipm.east);
        \draw[-, thin, black] (prnet.west) .. controls  +(south:10mm) and +(north:15mm)   .. (prm.north);
        \draw[-, thin, black] (prwa.south) .. controls  +(south:5mm) and +(north:1mm)   .. (prm.north);
        \draw[dashed, -, thin, red] (pnnet.south) .. controls  +(south:8mm) and +(south east:10mm)   .. (pnm.east);
    \end{tikzpicture}
\end{minipage}%
\vspace{.0001in}
\begin{minipage}{0.35\textwidth}
    \begin{tikzpicture}[
       AA/.style={circle, draw=black!60, fill=black!5, thick, minimum size=0.01cm, font=\tiny},
        BB/.style={circle, draw=red!60, fill=red!5, thick, minimum size=0.01cm, font=\tiny},
        CC/.style={circle, draw=blue!60, fill=blue!5, thick, minimum size=0.01cm, font=\tiny},
        DD/.style={circle, draw=green!60, fill=green!5, thick, minimum size=0.01cm, font=\tiny},
        EE/.style={circle, draw=yellow!60, fill=yellow!5, thick, minimum size=0.01cm, font=\tiny},
        FF/.style={circle, draw=orange!60, fill=orange!5, thick, minimum size=0.01cm, font=\tiny},
        GG/.style={circle, draw=cyan!60, fill=blue!5, thick, minimum size=0.01cm, font=\tiny},
        QQ/.style={circle, draw=purple!60, fill=purple!5, thick, minimum size=0.01cm, font=\tiny},
        ]
        \node at (-1.4, -3) [font=\tiny] {\tiny[{D}] Refined metadata relationships.};
        \node [AA] (net) {\textbf{nl}};
        \node[CC] (sipnet) [right=.1cm of net] {\tiny ip};
        \node[GG] (prnet) [below=.1cm of net] {\tiny pr};
        \node[BB] (tsnet) [above=.1cm of net] {\tiny t};
        \node[GG] (pnnet) [left=.1cm  of net] {\tiny pn};
        \draw[-, thin] (net.east) to (sipnet.west);
        \draw[-, thin] (net.north) to (tsnet.south);
        \draw[-, thin] (net.south) to (prnet.north);
        \draw[-, thin] (net.west) to (pnnet.east);
        \node [AA] (m)         [below left=1 of prnet]            {\textbf{m}};
        \node[BB] (tsm)        [above=.1 of m]                       {t};
        \node[DD]  (anm)       [right=.1 of m]                      {an}; 
        \node[GG] (prm)         [left=.1 of m]                       {pr};
        \node[CC] (sipm)       [below right=.3 of m]                      {ip};
        \node[GG] (pnm)       [below left=.3 of m]                      {pn};
        \node [AA] (wa)        [left=1.5 of net]            {\textbf{wa}};
        \node[BB] (twa)        [above=.1 of wa]                       {t};
        \node[DD] (anwa)        [left=.1 of wa]                       {an};
        \node[GG] (prwa)        [below=.1 of wa]                     {pr};
        \draw[-, thin] (net.east) to (sipnet.west);
        \draw[-, thin] (net.north) to (tsnet.south);
        \draw[-, thin] (net.south) to (prnet.north);
        \draw[-, thin] (net.west) to (pnnet.east);
        \draw[-, thin] (m.east) to (anm.west);
        \draw[-, thin] (m.north) to (tsm.south);
        \draw[-, thin] (m.south) to (sipm.north west);
        \draw[-, thin] (m.south) to (pnm.north east);
        \draw[-, thin] (m.west) to (prm.east);
        \draw[-, thin] (wa.north) to (twa.south);
        \draw[-, thin] (wa.south) to (prwa.north);
        \draw[-, thin] (wa.west) to (anwa.east);
        \draw[-, thin, black] (tsnet.west) .. controls  +(north:1mm) and +(north:1mm)   .. (twa.east);
       \draw[-, thin, black] (tsnet.west) .. controls  +(north west:5mm) and +(north:5mm)   .. (tsm.north);
        \draw[-, thin, black] (twa.east) .. controls  +(north east:1mm) and +(north:5mm)   .. (tsm.north);
        \draw[-, thin, black] (anwa.south) .. controls  +(south:2mm) and +(north:5mm)   .. (anm.north);
        \draw[-, thin, black] (sipnet.south) .. controls  +(south:2mm) and +(north:0mm)   .. (sipm.east);
        \draw[-, thin, black] (prnet.west) .. controls  +(south:10mm) and +(north:15mm)   .. (prm.north);
        \draw[-, thin, black] (prwa.south) .. controls  +(south:5mm) and +(north:1mm)   .. (prm.north);
    \end{tikzpicture}
\end{minipage}%
\begin{minipage}{0.35\textwidth}
    \begin{tikzpicture}[
        AA/.style={circle, draw=black!60, fill=black!5, thick, minimum size=0.01cm, font=\tiny},
         BB/.style={circle, draw=red!60, fill=red!5, thick, minimum size=0.01cm, font=\tiny},
         CC/.style={circle, draw=blue!60, fill=blue!5, thick, minimum size=0.01cm, font=\tiny},
         DD/.style={circle, draw=green!60, fill=green!5, thick, minimum size=0.01cm, font=\tiny},
         EE/.style={circle, draw=yellow!60, fill=yellow!5, thick, minimum size=0.01cm, font=\tiny},
         FF/.style={circle, draw=orange!60, fill=orange!5, thick, minimum size=0.01cm, font=\tiny},
         GG/.style={circle, draw=cyan!60, fill=blue!5, thick, minimum size=0.01cm, font=\tiny},
         QQ/.style={circle, draw=purple!60, fill=purple!5, thick, minimum size=0.01cm, font=\tiny},
         ]
        \node at (-1.4, -3) [font=\tiny] {\tiny[{E}] Evidence harmonisation through};
        \node at (-1.4, -3.2) [font=\tiny] {\tiny cross-matching timestamps.};
         \node [AA] (net) {\textbf{nl}};
         \node [AA] (net1) [below left=1.6 of net] {\textbf{nl}};
         \node[BB] (tsnet) [above=.1cm of net] {\tiny t1};
         \node[BB] (tsnet2)        [above=.1 of net1]                       {t2};
         \draw[-, thin] (net.north) to (tsnet.south);
         \node [AA] (m)         [below left=.6 of net]            {\textbf{m}};
         \node[BB] (tsm)        [above=.05 of m]                       {t1};
         \node [AA] (m1)        [below left=1.7 of m]            {\textbf{m}};
         \node[BB] (tsm1)        [above=.1 of m1]                       {t2};
         \node [AA] (wa)        [left=1 of net]            {\textbf{wa}};
         \node[BB] (twa)        [above=.1 of wa]                       {t1};
         \node [AA] (wa1)        [below left=1.5 of wa]            {\textbf{wa}};
         \node[BB] (twa1)        [above=.1 of wa1]                       {t2};
         \draw[-, thin] (net.north) to (tsnet.south);
         \draw[-, thin] (net1.north) to (tsnet2.south);
         \draw[-, thin] (m.north) to (tsm.south);
         \draw[-, thin] (m1.north) to (tsm1.south);
         \draw[-, thin] (wa.north) to (twa.south);
         \draw[-, thin] (wa1.north) to (twa1.south);
         \draw[-, thin, black] (tsnet.north) .. controls  +(north:2mm) and +(north:2mm)   .. (twa.north);
         \draw[-, thin, black] (tsnet.west) .. controls  +(north:5mm) and +(north:0mm)   .. (tsm.north);
         \draw[-, thin, black] (twa.east) .. controls  +(east:1mm) and +(north:5mm)   .. (tsm.north);
         \draw[-, thin, black] (tsnet2.north) .. controls  +(north:5mm) and +(north:5mm)   .. (twa1.north);
         \draw[-, thin, black] (tsnet2.west) .. controls  +(north:5mm) and +(north:0mm)   .. (tsm1.north);
         \draw[-, thin, black] (twa1.east) .. controls  +(east:1mm) and +(north:5mm)   .. (tsm1.north);
    \end{tikzpicture}
\end{minipage}%
\begin{minipage}{0.35\textwidth}
     \begin{tikzpicture}[
         AA/.style={circle, draw=black!60, fill=black!5, thick, minimum size=0.01cm, font=\tiny},
         BB/.style={circle, draw=red!60, fill=red!5, thick, minimum size=0.01cm, font=\tiny},
         CC/.style={circle, draw=blue!60, fill=blue!5, thick, minimum size=0.01cm, font=\tiny},
         DD/.style={circle, draw=green!60, fill=green!5, thick, minimum size=0.01cm, font=\tiny},
         EE/.style={circle, draw=yellow!60, fill=yellow!5, thick, minimum size=0.01cm, font=\tiny},
         FF/.style={circle, draw=orange!60, fill=orange!5, thick, minimum size=0.01cm, font=\tiny},
         GG/.style={circle, draw=cyan!60, fill=blue!5, thick, minimum size=0.01cm, font=\tiny},
         QQ/.style={circle, draw=purple!60, fill=purple!5, thick, minimum size=0.01cm, font=\tiny},
         ]
         \node at (-1.4, -3) [font=\tiny] {\tiny[{F}] Evidence harmonisation through};\label{fig:subfig:a}
         \node at (-1.4, -3.2) [font=\tiny] {\tiny  cross-matching IPs.}; 
         \node [AA] (net) {\textbf{nl}};
         \node [AA] (net1) [below left=1.6 of net] {\textbf{nl}};
         \node[CC] (tsnet) [above=.1cm of net] {\tiny ip1};
         \node[CC] (tsnet2)        [above=.1 of net1]                       {ip2};
         \draw[-, thin] (net.north) to (tsnet.south);
         \node [AA] (m)         [below left=.6 of net]            {\textbf{m}};
         \node[CC] (tsm)        [above=.05 of m]                  {ip1};
         \node [AA] (m1)        [below left=1.7 of m]             {\textbf{m}};
         \node[CC] (tsm1)       [above=.1 of m1]                  {ip2};
         \draw[-, thin] (net.north) to (tsnet.south);
         \draw[-, thin] (net1.north) to (tsnet2.south);
         \draw[-, thin] (m.north) to (tsm.south);
         \draw[-, thin] (m1.north) to (tsm1.south);
         \draw[-, thin, black] (tsnet.west) .. controls  +(north:3mm) and +(north:0mm)   .. (tsm.north);
         \draw[-, thin, black] (tsnet2.west) .. controls  +(north:5mm) and +(north:0mm)   .. (tsm1.north);
    \end{tikzpicture}
\end{minipage}%
\vspace{.0001in}

\vspace{1pt} 

\rule{\linewidth}{0.4pt} 

\vspace{1pt} 

\begin{minipage}{1\textwidth}
    \begin{tikzpicture}[
        AA/.style={circle, draw=black!60, fill=black!5, thick, minimum size=0.01cm, font=\tiny},
        BB/.style={circle, draw=red!60, fill=red!5, thick, minimum size=0.01cm, font=\tiny},
        CC/.style={circle, draw=blue!60, fill=blue!5, thick, minimum size=0.01cm, font=\tiny},
        DD/.style={circle, draw=green!60, fill=green!5, thick, minimum size=0.01cm, font=\tiny},
        EE/.style={circle, draw=yellow!60, fill=yellow!5, thick, minimum size=0.01cm, font=\tiny},
        FF/.style={circle, draw=orange!60, fill=orange!5, thick, minimum size=0.01cm, font=\tiny},
        GG/.style={circle, draw=cyan!60, fill=blue!5, thick, minimum size=0.01cm, font=\tiny},
        QQ/.style={circle, draw=purple!60, fill=purple!5, thick, minimum size=0.01cm, font=\tiny},
        PP/.style={circle, draw=black!0, fill=black!0, thick, minimum size=0.15cm, font=\fontsize{2mm}{2mm}\selectfont},
    ]
    \node [AA, minimum size=0.009cm, inner sep=2pt] (net) at (0, 0) {};
    \node [font=\tiny, right=0.0009cm of net] {\tiny nl:\tiny Network~Log};

    \node [AA, minimum size=0.009cm, inner sep=2pt, right=1.9 of net] (m) {};
    \node [font=\tiny, right=0.009cm of m] {m:Memory};

    \node [AA, minimum size=0.009cm, inner sep=2pt, right=1.5 of m] (wa) {};
    \node [font=\tiny, right=0.009cm of wa] {\tiny wa:Web Application};

    \node [CC, minimum size=0.009cm, inner sep=2pt, below=0.009cm of net] (ip) {};
    \node [font=\tiny, right=0.009cm of ip] {\tiny ip:IP Address};
    
    \node [BB, minimum size=0.009cm, inner sep=2pt, below=0.009cm of m] (time) {};
    \node [font=\tiny, right=0.009cm of time] {\tiny t:Timestamp};
    
    \node [GG, minimum size=0.009cm, inner sep=2pt, below=0.009cm of wa] (port) {};
    \node [font=\tiny, right=0.009cm of port] {\tiny pn:Port Number};
    
    \node [GG, minimum size=0.009cm, inner sep=2pt, right=2.4 of wa] (protocol) {};
    \node [font=\tiny, right=0.009cm of protocol] {\tiny pr:Protocol};
    
    \node [DD, minimum size=0.009cm, inner sep=2pt, right=1.4cm of protocol] (process) {};
    \node [font=\tiny, right=0.009cm of process] {\tiny an:\tiny Application Name};
    
    \node [QQ, minimum size=0.009cm, inner sep=2pt, right=2.4 of port] (url) {};
    \node [font=\tiny, right=0.009cm of url] {\tiny url:Uniform Resource Locator};
     
     \draw ([xshift=-5pt]net.north west) -- ++(-0, -12pt) node[midway, left=0pt] {\tiny Graph Keys};
    \end{tikzpicture}
\end{minipage}
\rule{\linewidth}{0.4pt} 
\vspace{-16pt} 
\caption{An exemplary depiction of the metadata harmonisation process.}
\label{fig:fig2}
\end{figure}

The process of mapping, validating, and enriching metadata involves applying various transformations and enhancements to the extracted metadata. This metadata mapping process includes mapping, refining, and enriching the represented nodes and metadata attributes to establish evidence relationships. As a result, it establishes connections and relationships among metadata entities, capturing dependencies, associations, and interdependencies among different metadata attributes. To illustrate the metadata harmonisation process descriptively, Fig.~\ref{fig:fig2} presents six subfigures. These subfigures depict the interactive workflow of initiating, refining, and querying metadata relationships among various entities, such as memory artefacts, network logs, and web browser data. The process begins with the unified system retrieving and defining metadata attributes within the data graph structure, as shown in sub-figure 'A' of Fig.~\ref{fig:fig2}. Subsequently, sub-figure 'B' illustrates the process of establishing cross-relationships among evidence entities based on their metadata types. During this stage, investigators interactively draw dashed black lines to establish the initial cross-matching between instances. The metadata initiated in sub-figure 'C' is then made available for investigators to review, edit, reject, or confirm. In this example, the investigative team rejects unnecessary nodes, such as the 'URL' node highlighted in the red dashed circle. They also reject the red dashed line linking two port numbers of entities associated with 'net' and 'm' nodes, as it is not relevant to the investigation. Subsequently, the outcome of the interactive engagement with the retrieved data modules is depicted in sub-figure 'D'. Lastly, the last two subfigures, sub-figure 'E' and sub-figure 'F' illustrate the outcomes of the cross-evidence query feature. For instance, the investigative team executes two queries based on cross-matching evidence timestamps and cross-matching IP addresses of entities, respectively.

\begin{figure}[htbp]
\centering
    \begin{minipage}{1\textwidth}
    \begin{tikzpicture}[
        AA/.style={circle, draw=black!60, fill=black!5, thick, minimum size=0.15cm, font=\fontsize{2mm}{2mm}\selectfont},
        BB/.style={circle, draw=red!60, fill=red!5, thick, minimum size=0.15cm, font=\fontsize{2mm}{2mm}\selectfont},
        CC/.style={circle, draw=blue!60, fill=blue!5, thick, minimum size=0.15cm, font=\fontsize{2mm}{2mm}\selectfont},
        DD/.style={circle, draw=green!60, fill=green!5, thick, minimum size=0.15cm, font=\fontsize{2mm}{2mm}\selectfont},
        EE/.style={circle, draw=yellow!60, fill=yellow!5, thick, minimum size=0.15cm, font=\fontsize{2mm}{2mm}\selectfont},
        FF/.style={circle, draw=orange!60, fill=orange!5, thick, minimum size=0.15cm, font=\fontsize{2mm}{2mm}\selectfont},
        GG/.style={circle, draw=cyan!60, fill=cyan!5, thick, minimum size=0.15cm, font=\fontsize{2mm}{2mm}\selectfont},
        PP/.style={circle, draw=black!0, fill=black!0, thick, minimum size=0.15cm, font=\fontsize{2mm}{2mm}\selectfont},
        ]
        \node [AA] (net)                                           {n};
        \node[CC] (sipnet)        [left =.5 of net]               {$sip$};
        \node[CC] (dipnet)        [right=.5 of net]               {$dip$};
        \node[FF] (sznet)         [below=.5 of net, label={[font=\scriptsize]below :4.2 GB}]          {$s$};
        \node[BB] (tsnet) [above=.5 of net,
                        label={[font=\scriptsize, label distance=.3cm]above:18/05/2022},
                                  label={[font=\scriptsize]above:13:55:12}] {$t$};
        \node [AA] (ec)           [right =1.5 of net]              {ec};
        \node [CC] (dipec)       [right=.1 of ec, label={[font=\scriptsize]above:93.125.188.22}]                  {$ip$};
        \node [AA] (dc) [left =2.5 of net] {dc};
        \node [BB] (tdc) [above =.3 of dc, 
                        label={[font=\scriptsize, label distance=.2cm]above:18/05/2022},
                        label={[font=\scriptsize]above:13:55:12}] {t$2$};
        \node [BB] (t2dc) [below=.3 of dc,label={[font=\scriptsize]below:10:15:15}] {t$1$};
        \node [CC] (ipdc) [right=.3 of dc] {$ip$};
        \node[above=0.1cm, font=\tiny] at ([xshift=6pt,yshift=43pt]dc.north) {\scriptsize \textcolor{red}{[A]}};
        \draw[dashed, red!50, thick, rounded corners] ([xshift=-16pt,yshift=50pt]dc.north west) rectangle ([xshift=4pt,yshift=-40pt]ipdc.south east);

        \draw[-, thin] (dc.east) to  (ipdc.west);
        \draw[-, thin] (dc.north) to  (tdc.south);
        \draw[-, thin] (dc.south) to  (t2dc.north);
        \node [AA] (sl)        [left =3 of dc]                   {slog};
        \node[BB] (tsl)        [above =.3 of sl,label={[font=\scriptsize, label distance=.3cm]above:18/05/2022},
                 label={[font=\scriptsize]above:13:55:12}] {$t$};
        \node[CC] (uipsl)      [right= .3 of sl, label={[font=\scriptsize]right:10.0.0.20}]           {$sip$};
        \node[DD] (ansl)       [left=.3 of sl, label={[font=\scriptsize]left:Alex}]                   {$an$};
        \node[DD] (cnsl)       [below=.3 of sl, label={[font=\scriptsize]below:Alex}]                 {$cn$};
        \draw[-, thin] (sl.north) to  (tsl.south);
        \draw[-, thin] (sl.east)       to  (uipsl.west);
        \draw[-, thin] (sl.west)       to  (ansl.east);
        \draw[-, thin] (sl.south)      to  (cnsl.north);
        \node [AA] (st)          [below= 2.6 of net]                   {s$1$};
        \node[BB] (tsst)         [above=.3 of st, 
                                                  label={[font=\scriptsize] right:18/05/2022},
                                                  label={[font=\scriptsize]above right:10:17:20}] {$t$};
        \node[CC] (sipst)        [right=.3 of st, label={[font=\scriptsize]right:10.0.0.20}]           {$ip$};
        \node[DD] (unst)         [left=.3 of st, label={[font=\scriptsize]left:Alex}]                  {$un$};
        \node [AA] (ai)        [left =4.5 of st]                    {ai};
        \node[FF] (szai)       [above=.3 of ai, label={[font=\scriptsize]above:4.2 GB}]                {$s$};
        \node[BB] (ctsai)      [left=.3 of ai]             {$t$};
        \node[DD] (anai)       [right=.3 of ai, label={[font=\scriptsize]right:Alex}]                  {$an$};
        \draw[-, thin] (net.west) to          (sipnet.east);
        \draw[-, thin] (net.east) to          (dipnet.west);
        \draw[-, thin] (net.north) to         (tsnet.south);
        \draw[-, thin] (net.south) to         (sznet.north);
         \draw[-, thin] (ec.east) to          (dipec.west);
        \draw[-, thin] (ai.north) to          (szai.south);
        \draw[-, thin] (ai.west) to           (ctsai.east);
        \draw[-, thin] (ai.east) to           (anai.west);
        \draw[-, thin] (st.north) to          (tsst.south);
        \draw[-, thin] (st.east) to           (sipst.west);
        \draw[-, thin] (st.west) to           (unst.east);
        \draw[-, thin, green!70!black] (sipst.north) .. controls +(west:20mm) and +(south:10mm) .. node[midway, above, text=green!70!black] {\scriptsize P} (uipsl.south);
        \draw[-, thin, green!70!black] (dipnet.south) .. controls +(south:8mm) and +(south:8mm) .. node[midway, above, text=green!70!black] {\scriptsize P} (dipec.south);
        \draw[-, thin, green!70!black] (sipnet.south) .. controls +(south:5mm) and +(south:5mm) .. node[midway, above, text=green!70!black] {\scriptsize P} (ipdc.south);
        \draw[-, thin, green!70!black] (tsnet.west) .. controls +(south:5mm) and +(south:5mm) .. node[midway, above, text=green!70!black] {\scriptsize P} (tdc.east);
        \draw[-, thin, green!70!black] (sznet.west) .. controls +(south:5mm) and +(east:5mm) .. node[midway, above, text=green!70!black] {\scriptsize P} (szai.east);
        \draw[-, thin, green!70!black] (unst.north) .. controls +(north:2mm) and +(north:2mm) .. node[midway, above, text=green!70!black] {\scriptsize P} (anai.north);
        \end{tikzpicture}
    \end{minipage}%
\vspace{0.0001in}
\vspace{0.0001 in}

\vspace{3pt} 

\rule{\linewidth}{0.4pt} 

\vspace{3pt} 

\begin{minipage}{1\textwidth}
    \begin{tikzpicture}[
        AA/.style={circle, draw=black!60, fill=black!5, thick, minimum size=0.01cm, font=\tiny},
        BB/.style={circle, draw=red!60, fill=red!5, thick, minimum size=0.01cm, font=\tiny},
        CC/.style={circle, draw=blue!60, fill=blue!5, thick, minimum size=0.01cm, font=\tiny},
        DD/.style={circle, draw=green!60, fill=green!5, thick, minimum size=0.01cm, font=\tiny},
        EE/.style={circle, draw=yellow!60, fill=yellow!5, thick, minimum size=0.01cm, font=\tiny},
        FF/.style={circle, draw=orange!60, fill=orange!5, thick, minimum size=0.01cm, font=\tiny},
        GG/.style={circle, draw=cyan!60, fill=blue!5, thick, minimum size=0.01cm, font=\tiny},
        QQ/.style={circle, draw=purple!60, fill=purple!5, thick, minimum size=0.01cm, font=\tiny},
        PP/.style={circle, draw=black!0, fill=black!0, thick, minimum size=0.15cm, font=\fontsize{2mm}{2mm}\selectfont},
    ]
      \node [AA, minimum size=0.009cm, inner sep=2pt] (net) at (0, 0) {};
      \node [font=\tiny, right=0.0009cm of net] {\tiny net:\tiny Network~Log};
      \node [AA, minimum size=0.009cm, inner sep=2pt, right=2.01 of net] (m) {};
      \node [font=\tiny, right=0 cm of m] {\tiny dc:\tiny Deleted~Cloud};
      \node [AA, minimum size=0.009cm, inner sep=2pt, right=1.95 of m] (slog) {};
      \node [font=\tiny, right=0 cm of slog] {slog:\tiny Syslog};
      \node [AA, minimum size=0.009cm, inner sep=2pt, below=.01 of net] (ec) {};
      \node [font=\tiny, right=0.0009cm of ec] {\tiny ec:\tiny External~Cloud};
      \node [AA, minimum size=0.009cm, inner sep=2pt, right=2.01 of ec] (ai) {};
      \node [font=\tiny, right=0 cm of ai] {\tiny ai:\tiny AI Model};
      \node [BB, minimum size=0.009cm, inner sep=2pt, right=1.3 of slog] (time) {};
      \node [font=\tiny, right=0 cm of time] {\tiny t:\tiny Time};
      \node [FF, minimum size=0.009cm, inner sep=2pt, right=0.9 of time] (s) {};
      \node [font=\tiny, right=0 cm of s] {\tiny s:\tiny Size};
      \node [DD, minimum size=0.009cm, inner sep=2pt, right=.7 of s] (un) {};
      \node [font=\tiny, right=0 cm of un] {\tiny u, c, a (n):\tiny Username};
      \node [CC, minimum size=0.009cm, inner sep=2pt, below=.01 of slog] (ip) {};
      \node [font=\tiny, right=0 cm of ip] {\tiny IP, dIP, sIP:\tiny source~or~destination~IP};
      \node [PP, minimum size=0.009cm, inner sep=2pt, right=3.93 of ip] (p) {};
      \node [font=\tiny, right=0cm of p, text=green!70!black] {\tiny P:\tiny Positive~match};
      \draw ([xshift=-5pt]net.north west) -- ++(-0, -12pt) node[midway, left=0pt] {\tiny Graph Keys};
  \end{tikzpicture}
\end{minipage} 
\rule{\linewidth}{0.4pt} 
\vspace{-16pt} 
\caption{A brief illustration of the metadata harmonisation case study results.}
\label{fig:fig3}
\end{figure}
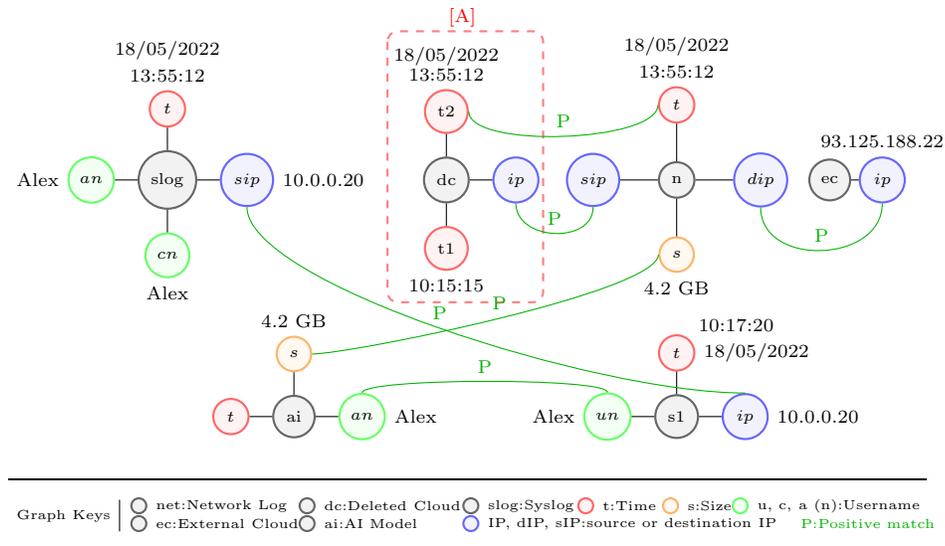

\noindent
Building upon the aforementioned process, Fig.~\ref{fig:fig3} illustrates the final result of the data harmonisation process, detailing potential evidence pertinent to this case study and addressing the first investigative question Q~\ref{Q1}. This visual representation implicates the suspect, Alex, in the leakage of the AI model. Distinct colours in the figure denote various attributes and entities: red nodes highlight timestamp events; orange nodes signify the AI model's size attribute; green nodes are events associated with suspect Alex; while light black nodes portray the primary data objects, illustrating the connections between network elements. Green lines mark positive matches between related entities. Notably, the red rectangle labelled with the initial 'A' pinpoints metadata attributes of the origin of the unidentified IP, addressing the second investigative question Q~\ref{Q2}. Overall, this figure elucidates the intricate interconnectedness and dependencies among different entities, offering a comprehensive view of the case study, and providing insight into the answer to investigative question Q~\ref{Q3}.

\subsubsection{Cross Analytics:} 
The proposed system's cross-analytics capabilities present a broad spectrum of analysis options. These options operate seamlessly across harmonised evidence, enhancing comprehensive digital forensics investigations. This study, however, emphasises the cross-timeline analysis to outline chronological events, thereby addressing Q~\ref{Q4} across the harmonised evidence demonstrated in the Fig.~\ref{fig:fig3}.

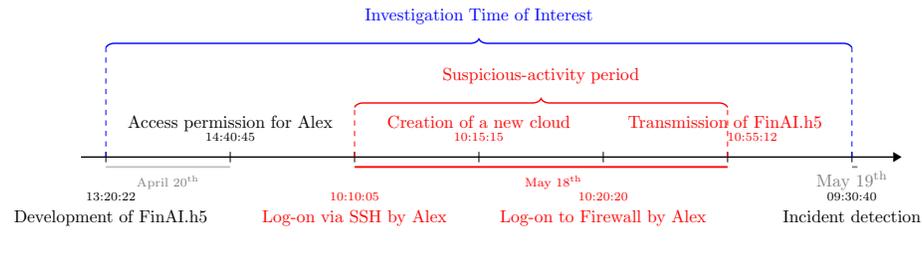
\begin{figure}[htbp]
    \centering
    \resizebox{1\textwidth}{!}{%
    \begin{tikzpicture}
            \draw[thick, -{Triangle}] (-.5,0) -- (16 cm,0) node[font=\scriptsize,below left=3pt and -8pt]{};
            \draw (0.1 cm, -17pt) node[anchor=north, font=\scriptsize]{13:20:22};
            \node[anchor=north, font=\normalsize, rotate=0] at (0.1 cm, -27pt) {Development of FinAI.h5};
            \draw (2.5 cm, +17pt) node[anchor=north, font=\scriptsize]{14:40:45};
            \draw (2.5 cm, +27pt) node[anchor=north, font=\normalsize]{Access permission for Alex};  
            \draw (5 cm, -17pt) node[anchor=north, font=\scriptsize] [red]{10:10:05};
            \draw (5 cm, -27pt) node[anchor=north, font=\normalsize] [red]{Log-on via SSH by Alex};  
            \draw (7.5 cm, +17pt) node[anchor=north, font=\scriptsize] [red]{10:15:15};
            \draw (7.5 cm, +27pt) node[anchor=north, font=\normalsize] [red]{Creation of a new cloud};      
            \draw (10 cm, -17pt) node[anchor=north, font=\scriptsize][red]{10:20:20};
            \draw (10 cm, -27pt) node[anchor=north, font=\normalsize][red]{Log-on to Firewall by Alex};    
            \draw (13 cm, +17pt) node[anchor=north, font=\scriptsize][red]{10:55:12};
            \draw (12.45 cm, +27pt) node[anchor=north, font=\normalsize][red]{Transmission of FinAI.h5};
            \draw (15 cm, -17pt) node[anchor=north, font=\scriptsize]{09:30:40};
            \draw (15 cm, -27pt) node[anchor=north, font=\normalsize]{Incident detection};
            \foreach \x in {0,2.5,...,15}
            \draw (\x cm,3pt) -- (\x cm,-3pt); 
            \foreach \x/\perccol in
            {0/100}
            \draw[lightgray!\perccol!gray, line width=1pt] 
            (\x,-.2) -- +(2.5,0);
            \node[anchor=north, font=\scriptsize, rotate=0] at (1.25cm, -07pt) [gray]{April 20\textsuperscript{th}};
            \foreach \x/\perccol in
            {5/80,7.5/0}
            \draw[white!\perccol!red, line width=1pt] 
            (5,-.2) -- +(7.5,0);
            \node[anchor=north, font=\scriptsize, rotate=0] at (9cm, -07pt) [red]{May 18\textsuperscript{th}};
            \foreach \x/\perccol in
            {15/0}
            \draw[white!\perccol!gray, line width=1pt] 
            (\x,-.2) -- +(0.1111,0);
            \node[anchor=north, font=\normalsize, rotate=0] at (15cm, -05pt) [gray]{May 19\textsuperscript{th}};
            [blue,midway,font=\textbf{}, above=10pt] {Investigation period};
            \draw [red][thick ,decorate,decoration={brace,amplitude=5pt}] (5,1)  -- +(7.5,0) 
                     node [black,midway,above=10pt, font=\normalsize] [red] {Suspicious-activity period};
            \draw[-, dashed, red] (5,0) --  +(0,1);
            \draw[-, dashed, red] (12.5,0) --  +(0,1);
            \draw [blue][thick ,decorate,decoration={brace,amplitude=5pt}] (0,+2.2)  -- +(15,0) 
                     node [blue,midway,font=\normalsize, above=10pt] {Investigation Time of Interest};
            \draw[-, dashed, blue] (0,0) --  +(0,2.2);
            \draw[-, dashed, blue] (15,0) --  +(0,2.2);
    \end{tikzpicture}%
}
\rule{\linewidth}{0.4pt} 
\vspace{-16pt} 
\caption{Timeline analysis of the harmonised timestamp events.}
\label{fig:fig4}
\end{figure}

\paragraph{Cross-timeline analysis:}
This form of analysis aims to construct a chronological representation of incident-related events. It aids in establishing timelines for activities and pinpointing relationships, dependencies, anomalies, and sequential temporal events. Pertaining to this case study, Table~\ref{tab:table6} delineates details of harmonised timeline events linked to the potential data of the case. It encapsulates data related to the evidence, focusing on specific aspects, especially events connected to the leaked 'FinAI.h5' model, as well as the origins of these timestamp events. In parallel, Fig.~\ref{fig:fig4} visually showcases these timeline events, tracing the entire life cycle of the 'FinAI.h5' model from its creation and deployment to the final moment of leakage.

\begin{table}[htbp]
\caption{Summary of the harmonised timestamp events.}
\label{tab:table6}
\resizebox{1\textwidth}{!}{%
\begin{tabular}{l|l|l|l|l|l|l|l}
\hline
Date       & Time     & Timestamp           & Category    & Type              & Attribute      & Value     & Metadata     \\
           &          & Attribute           &             &                   &                &           &  Source       \\ \hline
20/04/2022 & 13:20:22 & Created             & Development & AI model          & Model name     & FinAI.h5  & File system   \\ \hline
20/04/2022 & 14:40:45 & Created             & Security    & Access permission & Username       & Alex      & Cloud monitoring alert \\ \hline
18/05/2022 & 10:10:05 & Accessed            & Connection  & SSH               & Username       & Alex      & Memory        \\ \hline
18/05/2022 & 10:15:15 & Created             & Deployment  & Cloud server      & Username       & Alex      & Cloud  monitoring alert            \\ \hline
18/05/2022 & 10:20:20 & Created             & Security    & Logged in         & Username       & Alex      & Syslog server \\ \hline
18/05/2022 & 10:55:12 & Created             & Network     & Transmission      & IP             & 10.0.0.15 & Network logs  \\ \hline
\end{tabular}%
}
\end{table}

\noindent
The timeline analysis, as illustrated in Fig.~\ref{fig:fig4}, began with the creation and deployment of the 'FinAI.h5' model on AIxz's primary cloud server on April 20\textsuperscript{th}, 2022, at 13:20:22. Shortly after, at 14:40:45, the Cloud monitoring system logged a security event, granting full access permissions to an employee named Alex. On May 18\textsuperscript{th}, 2022, at 10:10:05, Alex leveraged his privileges and accessed the system via Secure Socket Shell (SSH). By 10:15:15, a new cloud server had been deployed, and the AI model was duplicated. At 10:20:20, a syslog event captured Alex accessing the perimeter firewall and establishing a new policy rule that allowed file transfer from the internal server 10.0.0.15 to external networks. The timeline analysis of the network traffic revealed that on May 18\textsuperscript{th}, 2022, at 11:33:10, a file approximately the size of the AI model was transferred to an external IP address. This in-depth timeline analysis offers insights into the sequence of events, highlighting potential security breaches and suspicious activities.

\section{Conclusion}\label{sec:con}
In the realm of digital forensics investigations, the harmonisation of heterogeneous evidence and the incorporation of advanced cross-data analytics are becoming paramount. This study proposes a novel approach to confront the challenges inherent to data heterogeneity and interoperability within digital evidence. The research findings underscore the  UMGM's efficacy in seamlessly integrating and harmonising diverse evidence data. A key feature of the system is its capability to capture intricate relationships of digital evidence across diverse formats. By unifying different evidence formats, it enables cross-data queries and facilitates cross-analysis across evidence. It also empowers investigators to navigate and refine interconnected evidence pieces, enhancing cross-platform interoperability. The conducted case study further validates the proposed approach and highlights its benefits. Through the unified approach, investigators could streamline their workflows, reduce manual efforts in correlating pieces of information, and improve the accuracy and speed of evidence analysis. Future research will focus on enhancing the UMGM's scalability for handling larger and more complex datasets. Additionally, efforts will be directed towards integrating machine learning algorithms into the system to automate advanced evidence analytics.


\begin{thebibliography}{99}

\bibitem{vincze2016challenges}
Eva A. Vincze,
\emph{Challenges in digital forensics},
\emph{Police Practice and Research},
vol. 17, no. 2, pp. 183--194, 2016,
Taylor \& Francis.

\bibitem{namjoshi2022role}
Janhavi Namjoshi and Manish Rawat,
\emph{Role of smart manufacturing in industry 4.0},
\emph{Materials Today: Proceedings},
vol. 63, pp. 475--478, 2022,
Elsevier.

\bibitem{casey2019chequered}
Eoghan Casey,
\emph{The chequered past and risky future of digital forensics},
\emph{Australian journal of forensic sciences},
vol. 51, no. 6, pp. 649--664, 2019,
Taylor \& Francis.

\bibitem{miller2023survey}
Christa M. Miller,
\emph{A survey of prosecutors and investigators using digital evidence: A starting point},
\emph{Forensic Science International: Synergy},
vol. 6, p. 100296, 2023,
Elsevier.

\bibitem{lillis2016current}
David Lillis, Brett Becker, Tadhg O'Sullivan, and Mark Scanlon,
\textit{Current challenges and future research areas for digital forensic investigation},
arXiv preprint arXiv:1604.03850 (2016).

\bibitem{rahman2020comprehensive}
Hafizur Rahman and Md Iftekhar Hussain,
\emph{A comprehensive survey on semantic interoperability for Internet of Things: State-of-the-art and research challenges},
\emph{Transactions on Emerging Telecommunications Technologies},
vol. 31, no. 12, p. e3902, 2020,
Wiley Online Library.


\bibitem{mohammed2018automating}
Hussam Mohammed, Nathan Clarke, and Fudong Li,
\emph{Automating the harmonisation of heterogeneous data in digital forensics},
in \emph{ECCWS 2018, 17th European Conference on Cyber Warfare and Security},
pp. 299--306, 2018.

\bibitem{alshumrani2023unified}
Ali Alshumrani, Nathan Clarke, and Bogdan Ghita,
\emph{A Unified Forensics Analysis Approach to Digital Investigation},
2023,
Academic Conferences International Ltd.

\bibitem{casino2022research}
Fran Casino, Thomas K. Dasaklis, Georgios P. Spathoulas, Marios Anagnostopoulos, Amrita Ghosal, Istvan Borocz, Agusti Solanas, Mauro Conti, and Constantinos Patsakis,
\emph{Research trends, challenges, and emerging topics in digital forensics: A review of reviews},
\emph{IEEE Access},
vol. 10, pp. 25464--25493, 2022,
IEEE.


\bibitem{brady2015deso}
Owen Brady, Richard Overill, and Jeroen Keppens,
\emph{DESO: Addressing volume and variety in large-scale criminal cases},
\emph{Digital Investigation},
vol. 15, pp. 72--82, 2015,
Elsevier.

\bibitem{chabot2015ontology}
Yoan Chabot, Aurélie Bertaux, Christophe Nicolle, and Tahar Kechadi,
\emph{An ontology-based approach for the reconstruction and analysis of digital incidents timelines},
\emph{Digital Investigation},
vol. 15, pp. 83--100, 2015,
Elsevier.

\bibitem{casey2015leveraging}
Eoghan Casey, Greg Back, and Sean Barnum,
\emph{Leveraging CybOX™ to standardize representation and exchange of digital forensic information},
\emph{Digital Investigation},
vol. 12, pp. S102--S110, 2015,
Elsevier.

\bibitem{casey2018evolution}
Eoghan Casey, Sean Barnum, Ryan Griffith, Jonathan Snyder, Harm van Beek, and Alex Nelson,
\emph{The evolution of expressing and exchanging cyber-investigation information in a standardized form},
\emph{Handling and Exchanging Electronic Evidence Across Europe},
pp. 43--58, 2018,
Springer.



\bibitem{arshad2020formal}
Humaira Arshad, Aman Jantan, Gan Keng Hoon, and Isaac Oludare Abiodun,
\emph{Formal knowledge model for online social network forensics},
\emph{Computers \& security},
vol. 89, p. 101675, 2020,
Elsevier.


\bibitem{sikos2020knowledge}
Leslie F. Sikos,
\emph{Knowledge representation to support partially automated honeypot analysis based on Wireshark packet capture files},
in \emph{Intelligent Decision Technologies 2019: Proceedings of the 11th KES International Conference on Intelligent Decision Technologies (KES-IDT 2019), Volume 1},
pp. 345--351, 2020,
Springer.

\bibitem{chikul2021ontology}
Pavel Chikul, Hayretdin Bahsi, and Olaf Maennel,
\emph{An ontology engineering case study for advanced digital forensic analysis},
in \emph{International Conference on Model and Data Engineering},
pp. 67--74, 2021,
Springer.

\bibitem{noel2016cygraph}
Steven Noel, Eric Harley, Kam Him Tam, Michael Limiero, and Matthew Share,
\emph{CyGraph: graph-based analytics and visualisation for cybersecurity},
in \emph{Handbook of Statistics},
vol. 35, pp. 117--167, 2016,
Elsevier.


\bibitem{schelkoph2019digital}
Daniel J. Schelkoph, Gilbert L. Peterson, and James S. Okolica,
\emph{Digital forensics event graph reconstruction},
in \emph{Digital Forensics and Cyber Crime: 10th International EAI Conference, ICDF2C 2018, New Orleans, LA, USA, September 10--12, 2018, Proceedings 10},
pp. 185--203, 2019,
Springer.


\end{thebibliography}
\end{document}